# COSMIC EMISSIVITY AND BACKGROUND INTENSITY FROM DAMPED LYMAN-ALPHA GALAXIES

S. Michael Fall

Space Telescope Science Institute, 3700 San Martin Drive, Baltimore, MD 21218

Stéphane Charlot

Institut d'Astrophysique du CNRS, 98 bis Boulevard Arago, 75014 Paris, France

AND

Yichuan C. Pei

Department of Physics and Astronomy, The Johns Hopkins University, Baltimore, MD 21218

## ABSTRACT

We present a new method to compute the cosmic emissivity $\mathcal{E}_\nu$ and background intensity $J_\nu$. Our method is based entirely on data from quasar absorption-line studies, namely, the comoving density of HI and the mean metallicity and dust-to-gas ratio in damped Ly$\alpha$ galaxies. These observations, when combined with models of cosmic chemical evolution, are sufficient to determine the comoving rate of star formation as a function of redshift. From this, we compute $\mathcal{E}_\nu$ and $J_\nu$ using stellar population synthesis models. Our method includes a self-consistent treatment of the absorption and reradiation of starlight by dust. In all of our calculations, the near-UV emissivity declines rapidly between $z \approx 1$ and $z = 0$, in agreement with estimates from the Canada-France Redshift Survey. The background intensity is consistent with a wide variety of observational limits and with a tentative detection at far-IR wavelengths.

*Subject headings:* cosmology: diffuse radiation — galaxies: evolution — quasars: absorption lines

## 1. INTRODUCTION

The mean emissivity of the universe $\mathcal{E}_\nu$ and the mean intensity of background radiation $J_\nu$ are important cosmological probes. The former is defined here as the power radiated per unit frequency per unit comoving volume, while the latter is defined as the power received per unit frequency per unit area of detector per unit solid angle of sky. These quantities, of course, are not independent; $J_\nu$ is given by an integral of $\mathcal{E}_\nu$ over redshift. It is likely that $\mathcal{E}_\nu$ and $J_\nu$ are dominated at near-UV, optical, and near-IR wavelengths by the direct radiation from stars and at far-IR wavelengths by reradiated starlight from the dust within galaxies. Thus, they contain potentially valuable information about the global history of star formation. Observationally, $\mathcal{E}_\nu$ and $J_\nu$ have proven to be elusive; until recently, it was only possible to estimate $\mathcal{E}_\nu$ at $z \ll 1$ and to place weak constraints on $J_\nu$. Theoretically, they have also proven to be elusive. Most calculations of $\mathcal{E}_\nu$ and $J_\nu$ are based on the properties of present-day galaxies and some assumed evolution in the past, often specified by several free parameters (see Lonsdale 1995 for a review).

In this Letter, we introduce a new method to compute $\mathcal{E}_\nu$ and $J_\nu$; in essence, we predict the "emission history" of the universe from its "absorption history." From the absorption lines in the spectra of distant quasars, it is possible in principle to determine the global rates of gas consumption, metal production, and hence star formation in galaxies at redshifts up to $z \approx 4$. The focus here is on the damped Ly$\alpha$ galaxies, which contain most of the cool, neutral gas in the universe and appear to be the progenitors of present-day galaxies (Lanzetta, Wolfe, & Turnshek 1995). We have already combined absorption-line data with models of cosmic chemical evolution to compute the comoving densities of stars and dust in damped Ly$\alpha$ galaxies (Pei & Fall 1995). Here, we employ stellar population synthesis models to compute $\mathcal{E}_\nu$ and $J_\nu$, including the absorption and reradiation of starlight by dust. In this first illustration of the method, we have deliberately kept the analysis as simple as possible in order to highlight the main ideas and assumptions. Moreover, the existing data on damped Ly$\alpha$ galaxies are too sparse to warrant a more elaborate analysis.

## 2. MODELS

Our first task is to relate the cosmic emissivity and background intensity, $\mathcal{E}_\nu$ and $J_\nu$, to the comoving densities of stars and dust, $\Omega_s$ and $\Omega_d$ (expressed here in units of the present critical density, $\rho_c = 3H_0^2/8\pi G$). We ignore other sources of radiation (primarily active galactic nuclei) and exclude photons that are absorbed in the galaxies in which they were produced. Thus, for the stellar part of the emissivity, we write

$$\mathcal{E}_{s\nu} = (1 - A_\nu)\mathcal{E}_{s\nu}^0, \qquad (1)$$

where $A_\nu$ is the mean fraction of absorbed photons, and $\mathcal{E}_{s\nu}^0$ is the emissivity before absorption. The latter is given by

$$\mathcal{E}_{s\nu}^0(t) = \rho_c \int_0^t dt' S_\nu(t-t') \frac{\dot{\Omega}_s(t')}{1-R(t')}, \qquad (2)$$

where $S_\nu(\Delta t)$ is the power radiated per unit frequency per unit initial mass by a generation of stars with an age $\Delta t$, $R$ is the returned fraction, and the dot denotes differentiation with respect to cosmic time. We compute $S_\nu$ and $R$ using the latest version of the Bruzual-Charlot models (described by Charlot, Worthey, & Bressan 1996). In most of the calculations reported here, the stellar initial mass function (IMF) is assumed to be a power law, $\phi(m) \propto m^{-(1+x)}$, with $x = 1.5$ and upper and lower cutoffs at $100 M_\odot$ and $0.1 M_\odot$. We further assume that all H ionizing photons are absorbed in the local interstellar medium, that 68% of them are converted to Ly$\alpha$ photons (the fraction appropriate for case B recombination in gas at $10^4$ K), that all of these are absorbed by dust (as a consequence of resonant scattering by HI), and that the remaining energy is radiated uniformly in wavelength between 3000 and 7000 Å (a range that includes most of the relevant emission lines). This treatment of ionizing radiation is consistent with far-UV observations of starburst galaxies (Leitherer et al. 1995).

We relate the mean fraction of photons absorbed by dust to the other properties of galaxies as follows. For simplicity, we assume that the stars and dust have the same spatial distributions within galaxies (so that the source function is constant along any ray), and we ignore the influence of scattering on absorption (which should be a good approximation for average quantities such as $A_\nu$). In this case, the fraction of photons absorbed along a ray that intercepts a galaxy with an optical depth $\tau_\nu$ is given by

$$a(\tau_\nu) = 1 - \tau_\nu^{-1}[1 - \exp(-\tau_\nu)]. \qquad (3)$$

We now define $\phi(\tau_\nu)d\tau_\nu$ to be the fraction of such rays with optical depths between $\tau_\nu$ and $\tau_\nu + d\tau_\nu$ when all galaxies and all positions and directions within them are considered. This enables us to express the mean fraction of photons absorbed by dust in the form

$$A_\nu = \frac{\int_0^\infty d\tau_\nu \tau_\nu \phi(\tau_\nu) a(\tau_\nu)}{\int_0^\infty d\tau_\nu \tau_\nu \phi(\tau_\nu)}. \qquad (4)$$

The optical depth and HI column density $N$ along a ray are related by $\tau_\nu = \kappa_\nu k_m m_H N$, where $\kappa_\nu$ is the opacity (i.e., mass absorption coefficient), $k_m$ is the dust-to-HI mass ratio, and $m_H$ is the mass of an H atom. Thus, for a single value of $k_m$, the (true) distributions of $\tau_\nu$ and $N$ are related by $\phi(\tau_\nu) \propto f(N)$. For consistency with the models of chemical evolution, we adopt the gamma distribution, $f(N) = (f_*/N)\exp(-N/N_*)$ (Pei & Fall 1995). This implies $\phi(\tau_\nu) \propto \tau_\nu^{-1}\exp(-\tau_\nu/\tau_{*\nu})$ with $\tau_{*\nu} = \kappa_\nu k_m m_H N_*$. Equations (3) and (4) and the relations $\Omega_d = k_m \Omega_{\rm HI}$ and $\Omega_{\rm HI} = (8\pi G m_H/3cH_0)f_* N_*$ then give

$$A_\nu = 1 - \tau_{*\nu}^{-1}\ln(1+\tau_{*\nu}), \qquad (5)$$

$$\tau_{*\nu} = (3cH_0/8\pi G f_*)\kappa_\nu \Omega_d. \qquad (6)$$

If the stars and dust have different spatial distributions or there is a dispersion in the dust-to-gas ratio, these expressions should be regarded as approximations. We have checked that our results are not sensitive to the particular form of $A_\nu$.

The starlight absorbed by dust is reradiated thermally. We approximate the corresponding emissivity by that of a single blackbody:

$$\mathcal{E}_{d\nu} = 4\pi \rho_c \Omega_d \kappa_\nu B_\nu(T_d). \qquad (7)$$

The effective temperature of the dust $T_d$ is then determined by the condition of energy balance

$$4\pi \rho_c \Omega_d \int_0^\infty d\nu \kappa_\nu [B_\nu(T_d) - B_\nu(T_{\rm CMB})] = \int_0^\infty d\nu A_\nu \mathcal{E}_{s\nu}^0, \qquad (8)$$

where $T_{\rm CMB} = 2.73(1+z)$K is the temperature of the cosmic microwave background radiation. Equation (7) should be a good approximation at wavelengths beyond the peak [$\lambda_{\max} \approx 150(T_d/20\,{\rm K})^{-1}$ $\mu$m], where the emission is likely to be dominated by large amounts of cool dust, but a poor approximation at shorter wavelengths, where the emission is likely to be dominated by small amounts of warm dust (e.g., small, transiently heated grains). We return to this point later. For the optical properties of the dust, we adopt the Draine & Lee (1984) model but with the proportions of graphite and silicates adjusted so as to reproduce the mean UV extinction curve in the Large Magellanic Cloud (Pei 1992). In this model, the opacity at long wavelengths has the form $\kappa_\nu \propto \nu^2$, and the left-hand side of equation (8) is proportional to $\Omega_d(T_d^6 - T_{\rm CMB}^6)$. Given $\Omega_s$ and $\Omega_d$, we can now compute the emissivity $\mathcal{E}_\nu = \mathcal{E}_{s\nu} + \mathcal{E}_{d\nu}$ from the equations above, and the corresponding background intensity at $z = 0$ from

$$J_\nu = \frac{c}{4\pi} \int_0^\infty dz \mathcal{E}_{(1+z)\nu} \left| \frac{dt}{dz} \right|. \quad (9)$$

Here, we have neglected absorption between the sources of radiation and the observer. This is appropriate because ionizing photons are assumed to be absorbed locally and because relatively few non-ionizing photons are absorbed by the dust in intervening galaxies.

The comoving densities of stars and gas in galaxies, $\Omega_s$ and $\Omega_g$, and the mean metallicity in the interstellar medium $Z$, including dust, are governed by the equations of cosmic chemical evolution:

$$\dot{\Omega}_g + \dot{\Omega}_s = \dot{\Omega}_f, \quad (10)$$

$$\Omega_g \dot{Z} - y\dot{\Omega}_s = (Z_f - Z)\dot{\Omega}_f. \quad (11)$$

Here, $y$ is the IMF-averaged yield, and the source terms on the right represent the inflow or outflow of gas with metallicity $Z_f$ at a rate $\dot{\Omega}_f$. For purposes of illustration, we assume that just over half of the heavy elements are locked up in dust grains and that the ionized and molecular components of the interstellar medium are negligible, i.e., $\Omega_d = 0.55 Z \Omega_g$ and $\Omega_g = 1.3 \Omega_{\rm HI}$. The constancy of the dust-to-metals ratio can also be expressed in the form $k(z)/Z(z) = k(0)/Z(0) = 0.8/Z_\odot$, where $k$ is the ratio of $\tau_B$, the extinction optical depth in the rest-frame $B$ band, to $N/10^{21}$ cm$^{-2}$ (see Table 2 of Pei 1992). This agrees to within a factor of two with the dust and metal content of present-day galaxies and damped Ly$\alpha$ galaxies at $\overline{z} = 2.2$ (Pei, Fall, & Bechtold 1991; Pettini et al. 1994). Given $\Omega_{\rm HI}$ and some assumptions about $\dot{\Omega}_f$ and $Z_f$, it is now possible to solve equations (10) and (11) for $\Omega_s$ and $\Omega_d$. We adopt the solutions presented by Pei & Fall (1995). These are of three types: a closed-box model ($\dot{\Omega}_f = 0$), a model with inflow of metal-free gas ($\dot{\Omega}_f = +\nu\dot{\Omega}_s$, $Z_f = 0$), and a model with outflow of metal-enriched gas ($\dot{\Omega}_f = -\nu\dot{\Omega}_s$, $Z_f = Z$). The adjustable parameters are the initial comoving density of gas in galaxies $\Omega_{g\infty}$ and the relative inflow or outflow rate $\nu$. The models include self-consistent corrections for the damped Ly$\alpha$ galaxies that are missing from op-tically selected samples as a result of the obscuration of background quasars (Fall & Pei 1993]). The models were designed to reproduce the observed comoving density of HI in damped Ly$\alpha$ galaxies at $0 < z \lesssim 4$ (Lanzetta et al. 1995, Storrie-Lombardi, McMahon, & Irwin 1996). They also reproduce the observed mean metallicity $Z \approx 0.1 Z_\odot$ at $\overline{z} = 2.2$ (Pettini et al. 1994) and are consistent with the average properties of present-day galaxies. The results presented here, unless otherwise noted, are based on models with $\Omega_{g\infty} = 4 \times 10^{-3} h^{-1}$, $\nu = 0.5$, and $h = 0.5$, $q_0 = 0.5$, $\Lambda = 0$ (with $h \equiv H_0/100$ km s$^{-1}$ Mpc$^{-1}$). For comparison, we also display results from the analogous models without dust.

## 3. RESULTS

Figure 1 shows the evolution of $\mathcal{E}_\nu$ at rest-frame wavelengths of 2800 Å and 1.0 $\mu$m. The former is dominated by the light from young, massive stars and is thus nearly a direct indicator of current star formation, while the latter includes the light from a wide range of stellar types and thus reflects a combination of past and current star formation. The solid curves in Figure 1 rep-resent the closed-box (C), inflow (I), and outflow (O) models with dust, while the dashed curves represent the analogous models without dust. The data points with error bars are estimates by Lilly et al. (1996) from local samples of galaxies ($z = 0$) and from the Canada-France Redshift Survey ($z = 0.3, 0.6,$ and $0.9$). These include corrections for incompleteness at faint magnitudes. The observed emissivity at 2800 Å decreases by an order of magnitude between $z = 0.9$ and $z = 0$, indicating a similar decrease in the comoving rate of star formation. Evidently, the slopes of the predicted and observed $\mathcal{E}_\nu(z)$ relations agree to within the uncertainties, both at 2800 Å and 1.0 $\mu$m. The amplitudes match for the closed-box and outflow models with dust but not for the inflow model with dust. However, the predicted amplitudes are somewhat sensitive to the adopted IMF in the population synthesis models. If the IMF slope were increased to $x = 2.0$ or the lower cutoff were decreased to $0.01 M_\odot$, the predicted amplitudes would be reduced by factors of $2-3$. With this freedom to adjust the IMF, the emissivities in the inflow model with dust can be brought into rough agreement with the observed emissivities. We count this as a success because models with very different histories of star formation cannot be made consistent with the observations for any choice of the IMF. For example, models with $\Omega_{g\infty} \lesssim 1 \times 10^{-3} h^{-1}$ or $\Omega_{g\infty} \gtrsim 8 \times 10^{-3} h^{-1}$ have emissivities that decline too slowly or too rapidly.

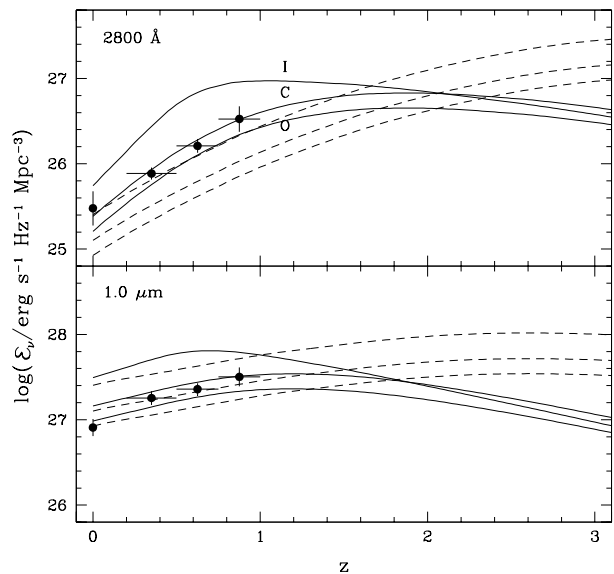

FIG. 1.—Cosmic emissivity $\mathcal{E}_\nu$ as a function of redshift $z$ at rest-frame wavelengths of 2800 Å (top) and 1.0 $\mu$m (bottom). The solid curves are from the closed-box (C), inflow (I), and outflow (O) models with dust, while dashed curves are from the analogous models without dust. The data points with error bars are estimates from local samples and the Canada-France Redshift Survey (Lilly et al. 1996).

Figure 2 shows $\nu J_\nu$ as a function of wavelength. Again, the solid curves represent the closed-box (C), inflow (I), and outflow (O) models with dust, while the dashed curves represent the analogous models without dust. The symbols with arrows and the hatched line represent the observational limits described in the caption. Evidently, our models are consistent with these constraints, over four and a half decades in wavelength. The near-UV background is higher in the models with dust because they have more star formation at low redshifts than the models without dust (see Figure 1 and Pei & Fall 1995). The open circles with vertical bars in Figure 2 represent the tentative detection of an extragalactic far-IR background by Puget et al. (1996). This result, derived from $COBE$/FIRAS data, is uncertain because it depends critically on the removal of foreground emission by interplanetary and interstellar dust. Our models with dust are consistent with this detection. The effective temperature of the dust remains near $T_d \approx 20$ K at $z \gtrsim 1$ and then decreases to $T_d \approx 15$ K at $z = 0$. This produces a peak in $\nu J_\nu$ at $\lambda \approx 240$ $\mu$m. The valley at $\lambda \approx 60$ $\mu$m is an artifact of our assumption that the spectrum of the dust emission is that of single blackbody [see equation (7)]. We have experimented with more realistic, two-temperature models and find that the valley can easily be filled in by emission from small amounts of warm dust with little effect on the emission at longer wavelengths. This indicates that the far-IR background is a robust feature of our models. The amplitude of $J_\nu$ depends on the IMF in the same way as the amplitude of $\mathcal{E}_\nu$; in particular, the background would be weaker at all wavelengths if the proportion of massive stars were reduced. Moreover, the values of $\Omega_{g\infty}$ that are consistent with the data on $J_\nu$ are essentially the same as those that are consistent with the data on $\mathcal{E}_\nu$.

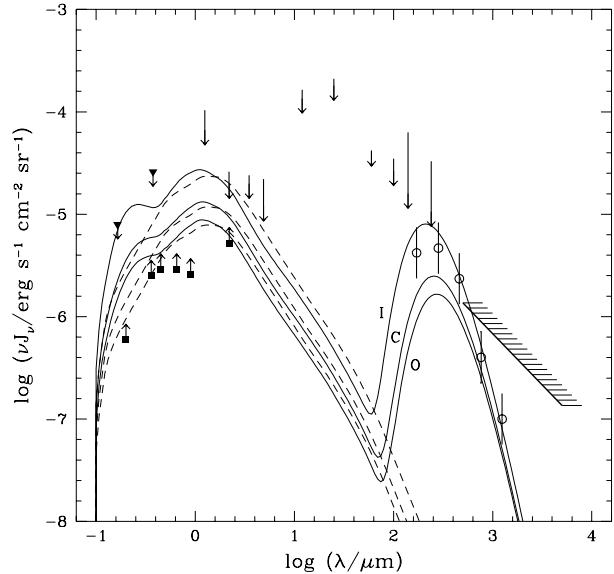

FIG. 2.—Background intensity $J_\nu$ at $z = 0$ times frequency $\nu$ as a function of wavelength $\lambda$. The solid curves are from the closed-box (C), inflow (I), and outflow (O) models with dust, while the dashed curves are from the analogous models without dust. The squares with arrows are lower limits derived from galaxy counts at 2000 Å (Milliard et al. 1992), 3600–9000 Å (Tyson 1995), and 2.2 $\mu$m (Cowie et al. 1994). The triangles with arrows are upper limits or possible detections at 1600 Å (Paresce 1990) and 4000 Å (Mattila 1990). The unadorned arrows are the residuals of $COBE$/DIRBE data after preliminary removal of foreground emission at 1.25–240 $\mu$m (Hauser 1996). The hatched line is a conservative upper limit based on $COBE$/FIRAS data at 500–5000 $\mu$m (twice the limit from Mather et al. 1994; see Hauser 1996). The open circles with vertical bars are a tentative detection derived from $COBE$/FIRAS data at 170–1260 $\mu$m (Puget et al. 1996).

## 4. DISCUSSION

We have computed the cosmic emissivity $\mathcal{E}_\nu$ and back-ground intensity $J_\nu$ with input only from absorption-line studies of damped Ly$\alpha$ galaxies. These objects have $N \gtrsim 10^{20}$ cm$^{-2}$ (by definition) and are the probable sites of most star formation in the universe. They appear to be the progenitors of present-day galaxies, although many of their properties remain to be determined. In particular, we do not yet know the sizes and morphologies of the damped Ly$\alpha$ galaxies. It is possible that most of those at low redshifts are disks, while most of those at high redshifts are spheroids. We emphasize that the results presented here are not affected by such issues. The reason for this is that all of the quantities required in our analysis can be computed directly from statistics of the absorption along random lines of sight [e.g., $\Omega_{\rm HI}$ is given by the integral of $Nf(N)$ over $N$]. The primary restriction on our results is that they do not include any galaxies that consumed their HI before $z \approx 4$, the highest redshift probed systematically by quasar absorption-line studies. Since our calculations include several approximations and idealizations, we regard them as illustrative rather than definitive. Nevertheless, we find reasonable agreement with all of the available data on $\mathcal{E}_\nu$ and $J_\nu$ without any fine-tuning of parameters. In particular, our models reproduce the rapid decline in the near-UV emissivity between $z \approx 1$ and $z = 0$ reported by Lilly et al. (1996). They are also consistent with a wide variety of observational limits on the extragalactic background and a tentative detection at far-IR wavelengths reported by Puget et al. (1996). It is therefore possible that this background is produced mainly or entirely by galaxies at $z \lesssim 4$.

We thank M.G. Hauser, N. Panagia, and the referee for helpful comments.